# Magnetic Field Line Stickiness in Tokamaks

Caroline G. L. Martins, M. Roberto and I. L. Caldas

**Abstract** - We present simulated figures of the diverted magnetic field lines of the tokamak ITER, obtained by numerically integrating a Hamiltonian model with electrical currents in five wire loops and control coils. We show evidences of a sticky island embedded in the chaotic region near the divertor plates, which traps magnetic field lines for many toroidal turns increasing their connection lengths to these plates.

The tokamak ITER, an international collaboration to test the magnetic confinement of plasmas for thermonuclear controlled fusion, will operate with a poloidal divertor. Thus, its magnetic configuration will contain magnetic field lines on nested toroidal surfaces and open field lines hitting the divertor plates. The brake of symmetry due to error fields, plasma oscillations and control fields will create chaotic field lines around the separatrix [1]. Moreover, the control of chaotic field lines that escape to the divertor plates will be essential to improve the plasma confinement.

In the present work we use a Hamiltonian model, based on the model introduced in [2], with five loop wires that reproduces the ITER-like magnetic surfaces topology, including a X-point related to the divertor. We add perturbations created by pairs of loop coils, similar to the field-error correction coils (C-coils) installed at the DIII-D tokamak [3] and those that will be installed in ITER.

The chaotic lines escape to the divertor plates, but some of them may be trapped, for many toroidal turns, in complex structures around magnetic islands, giving rise to the so called stickiness effect [4]. The numbers of toroidal turns performed by magnetic field lines until reaching the divertor plates are called connection lengths, and their distribution directly interfere on the heat flux transport, determining the deposition patterns on the divertor plates [3]. Then, it is essential to analyze the topology of sticky structures, and a way to do that is trough the numerical calculation of the finite time rotation number (FTRN) [5].

The FTRN is a recently proposed numerical diagnostic to identify different regions of stickiness [5]. The FTRN measures the average rotation angle performed by a field line over a chosen axis as a reference, in a finite number of iterations N (in our case for N = 205 toroidal turns). While the rotation number (in our case the inverse of the safety factor) is not well-defined for chaotic orbits, its FTRN value exists for any orbit, chaotic or not.

Figure 1(a) shows the Poincaré section for the system perturbed by ten pairs of coils with $I_c$ = 70 kA. The magnetic field lines near the X-point are chaotic and, eventually, end their trajectories on the divertor plates, located horizontally at Z = -3.7 m. Figure 1(b) shows the global FTRN for field lines with initial conditions in the red rectangle showed at (a), and it was calculated adopting the center of the plasma column (R ≈ 6.41 m, Z ≈ 0.513 m) as the reference for the field lines winding number. As well as the finite time Lyapunov exponent, the FTRN, when calculated forward, evidences the stable manifolds of the unstable periodic orbits. These manifolds act like barriers separating different regions of the chaotic sea [5], explaining the colorful parallel structures formed by small resonances in (b).

After a connection length analysis of the resonances near the X-point, we present an example of a strong stickiness at the border of the island located at the black rectangle in (b). In order to understand the mentioned effect, we calculate the local FTRN adopting the center of the sticky island (R ≈ 4.9474 m, Z ≈ -3.0710 m) as the reference for the field lines winding number. Figure (c) shows the local FTRN, in a logarithmic scale, and one can notice four escape channels (in orange) created by a stable manifold of an unstable periodic orbit in the outer border of the sticky island. Figure (d) shows the local FTRN, in a linear scale, emphasizing complex structures created by cantori surrounding the sticky island [5]. The cantori gaps are usually very small, thus a chaotic orbit inside the cantorus takes a long time before escaping to the outer chaotic sea, and a stickiness phenomenon appears [4].

The figures presented here indicate that the stickiness reported in this work reduces the field line transport to the divertor plates.

Manuscript received X XXX 2013; revised X XXX 2014.
Caroline G. L. Martins and M. Roberto with *Instituto Tecnológico de Aeronáutica, Departamento de Física, São José dos Campos - Brazil*. I. L. Caldas with *Universidade de São Paulo, Instituto de Física, São Paulo - Brazil*.
The authors would like to thank J. D. Szezech for useful comments, and the following Brazilian scientific agencies for the financial support: CAPES, CNPq and São Paulo Research Foundation (FAPESP), Grants 2010/13162-0, 2013/03401-6 and 2011/19269-11.
Publisher Identifier S XXXX-XXXXXXX-X

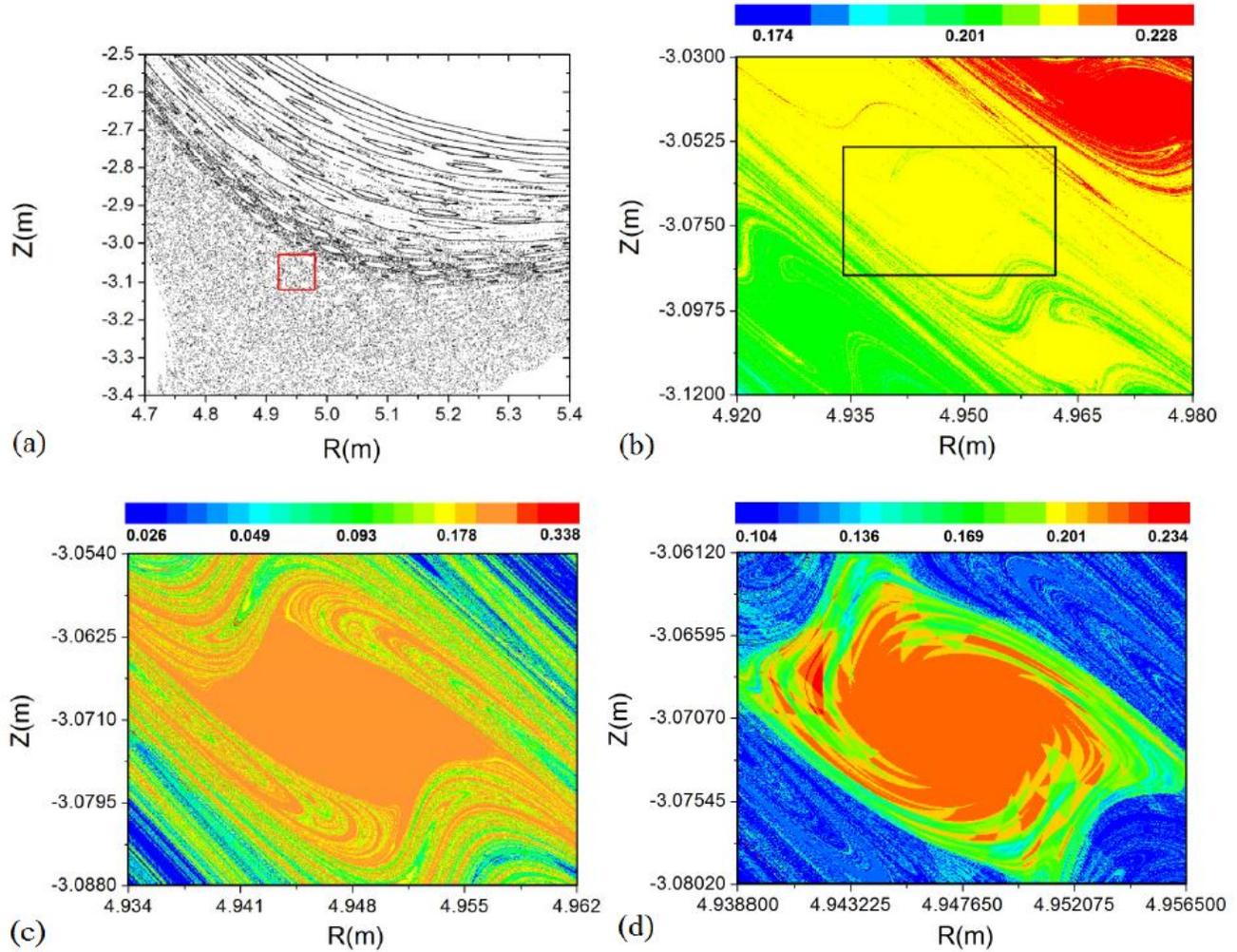

Figure 1 (a) Poincaré section for the Hamiltonian perturbed by ten pairs of coils with $I_c = 70$ kA. (b) Global FTRN for lines in the red rectangle of (a), calculated adopting the center of the plasma column (R ≈ 6.41 m, Z ≈ 0.513 m) as the reference for the field line winding number. (c) Local FTRN, in a logarithmic scale, calculated adopting the center of the sticky island (R ≈ 4.9474 m, Z ≈ -3.0710 m) as the reference for the field line winding number. (d) Local FTRN, in a linear scale, emphasizing complex structures created due to the existence of cantori surrounding the sticky island.

Manuscript received X XXX 2013; revised X XXX 2014.
Caroline G. L. Martins and M. Roberto with *Instituto Tecnológico de Aeronáutica, Departamento de Física, São José dos Campos - Brazil*. I. L. Caldas with *Universidade de São Paulo, Instituto de Física, São Paulo - Brazil*.
The authors would like to thank J. D. Szezech for useful comments, and the following Brazilian scientific agencies for the financial support: CAPES, CNPq and São Paulo Research Foundation (FAPESP), Grants 2010/13162-0, 2013/03401-6 and 2011/19269-11.
Publisher Identifier S XXXX-XXXXXXX-X